\documentclass[twoside]{ae100prg}
\usepackage[breaklinks]{hyperref}
\usepackage{booktabs}
\usepackage{amsfonts}
\usepackage{amsmath}
\usepackage{amssymb}
\usepackage{subfig}
\usepackage[dvips]{graphicx}

\newcommand{\x}{\mathsf{x}}
\newcommand{\X}{\mathsf{X}}

\DeclareMathOperator{\Realpart}{Re}

\begin{document}
\title{Unruh-DeWitt detector on the BTZ black hole\footnote{Based
on a talk given by L.H. at ``Relativity and Gravitation:
100 Years after Einstein in Prague" 
, Prague, 25 June -- 29 June 2012.}}


\author{Lee Hodgkinson$^1$, Jorma Louko$^{1,2}$}

\address{$^1$ School of Mathematical Sciences,
University of Nottingham, Nottingham NG7 2RD, UK}

\address{$^2$ Kavli Institute for Theoretical Physics, 
University of California, Santa Barbara, CA 93106-4030, USA}

\email{pmxlh1@nottingham.ac.uk, jorma.louko@nottingham.ac.uk}

\begin{abstract}
We examine an Unruh-DeWitt particle detector coupled to a scalar field
in three-dimensional curved spacetime, within first-order perturbation theory. 
We first obtain a causal and manifestly regular expression for the 
instantaneous transition rate in an
arbitrary Hadamard state. 
We then specialise to the Ba\~nados-Teitelboim-Zanelli black hole 
and to a massless conformally coupled field in the Hartle-Hawking vacuum. 
A~co-rotating detector responds thermally in the expected local Hawking temperature, 
while a freely-falling detector shows no evidence of thermality 
in regimes that we are able to probe, not even far from the horizon. 
The boundary condition at the asymptotically anti-de~Sitter 
infinity has a significant effect on
the transition rate. \vspace{-2ex}

\end{abstract}

\section{Introduction}
\label{sec:intro}

Whenever non-inertial observers or curved backgrounds are present in quantum field theory, 
the notions of vacuum state and particle number 
become non-unique. 
For this reason it proves convenient to define particles operationally;
that is to say, we couple the field to a simple quantum mechanical 
system that we think of as our detector and define particles via the field's
interaction with the energy levels of this system. Upwards (respectively downwards)
transitions can be interpreted as due to absorption (emission) 
of field quanta, or particles. This is the Unruh-DeWitt 
model for a particle detector~\cite{deWitt,unruh}. 

In this contribution we address 
a pointlike Unruh-DeWitt detector 
coupled to a scalar field in three-dimensional spacetime, within first-order perturbation theory. 
We first find the detector's instantaneous transition rate in an arbitrary Hadamard state. 
We then specialise to a massless conformally coupled field on the Ba\~ndados-Teitelboim-Zanelli (BTZ) black hole, 
in the Hartle-Hawking vacuum, analysing the thermal response seen by a co-rotating detector and the time evolution of the response of a freely-falling detector. A~longer exposition of the results 
can be found in~\cite{Hodgkinson:2012mr}.

\section{Transition rate in $(2+1)$ dimensions}
\label{sec:transitionrate}

With the Unruh-DeWitt detector, the fundamental quantity of interest is the probability of a transition between the energy eigenstates. In the framework of first order perturbation theory the probability for a transition of energy $E$ is proportional to the response function, 
\begin{equation}
\label{eq:respfunc}
\mathcal{F}(E)=2\,
\lim_{\epsilon\to0_+} 
\Realpart \int_{-\infty}^{\infty}\mathrm{d}u \,
\chi(u)\int_{0}^{\infty}\mathrm{d}s \,\chi(u-s)\, 
\mathrm{e}^{-iE s}\, W_\epsilon(u,u-s)\,,
\end{equation}
where $\chi$ is a smooth switching function that turns on (off) the detector's interaction with the field and  $W_\epsilon(u,u-s)$ is a one-parameter family of functions that converge to the pull-back of the Wightman distribution to the detector's wordline~\cite{Fewster:1999gj,junker,kay-wald,satz-louko:curved}.
A related quantity of interest is the transition rate, 
which can be defined as the derivative of the transition probability 
with respect to the total detection time. One must take great care when obtaining the transition rate from
the response function~\cite{Langlois,louko-satz:profile,Sriramkumar:1994pb,schlicht}. 
We will adopt the approach developed in~\cite{hodgkinson-louko:beyond4d,satz-louko:curved,satz:smooth} of taking a 
controlled sharp switching limit.

In three-dimensional spacetime, the Wightman distribution $W(\x,\x')$ 
of a real scalar field in a Hadamard state can be represented by the $\epsilon\to0_+$ limit of 
a family of functions with the short distance
form \cite{Decanini:2005gt}
\begin{equation}
\label{eq:3dHadamard}
W_{\epsilon}(\x,\x')=\frac{1}{4\pi}\left[\frac{U(\x,\x')}{\sqrt{\sigma_{\epsilon}(\x,\x')}}
+\frac{H(\x,\x')}{\sqrt{2}}\right] , 
\end{equation}
where $\epsilon$ is a positive parameter, 
$\sigma(\x,\x')$ is the squared geodesic distance between $\x$ and~$\x'$, 
$\sigma_{\epsilon}(\x,\x'):=\sigma(\x,\x')+2i\epsilon\left[T(\x)-T(\x')\right]+\epsilon^2$ 
and $T$ is any globally-defined future-increasing $C^{\infty}$ function. 
The branch of the square root is such that the $\epsilon\to 0_+$ 
limit of the square root is positive when 
$\sigma(\x,\x')>0$~\cite{Decanini:2005gt,kay-wald}. 
Here $U(\x,\x')$ and $H(\x,\x')$ are symmetric biscalars which 
have expansions governed by certain recursion relations~\cite{Decanini:2005gt}, and they 
are regular in the coincidence limit. 

Given~\eqref{eq:3dHadamard}, the detector's instantaneous 
transition rate can be shown to take the form \cite{Hodgkinson:2012mr}
\begin{align}
\label{eq:resp3d:sharp:rate}
\dot{\mathcal{F}}_{\tau}\left(E\right)
=
\frac{1}{4}+2\int^{\tau-\tau_0}_{0}\,\mathrm{d}s
\Realpart\left[\mathrm{e}^{-iEs} \, W_0(\tau,\tau-s)\right]
\ , 
\end{align}
where $\tau_0$ is the proper time at which the detector was switched
on, $\tau$ is the proper time at which the instantaneous transition
rate is read off, and the function $W_0$ is the pointwise
$\epsilon\to0_+$ limit of $W_\epsilon$.  We are assuming that any
singularities that $W(\x,\x')$ may have at $\sigma(\x,\x')\neq 0$, not
captured by the asymptotic
expansion~\eqref{eq:3dHadamard}, are so mild that taking the pointwise
limit is valid.  Such singularities will in particular occur in the
BTZ spacetime below.

\section{Detector in the BTZ spacetime}
\label{sec:BTZoverview}

\begin{figure}[h!]
\center
\includegraphics[width=0.35\textwidth]{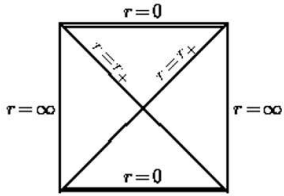}
\caption{Conformal diagram for the $J=0$ BTZ black hole. 
The Killing horizon of the Killing vector $\partial_t$ is at $r=r_+$, where 
$r_+ = \sqrt{M}\,\ell$.
\label{fig:BTZstatic}}
\end{figure}

We now specialise to a detector in the BTZ black hole spacetime
\cite{HBTZ,BTZ,Carlip:1995}. This spacetime can be obtained by
periodically identifying $\text{AdS}_3$, and in coordinates adapted to
the global isometries the metric takes the form
\begin{equation}
ds^2 = -( N^\perp)^2dt^2 + f^{-2}dr^2 + r^2\left( d\phi + N^\phi dt\right)^2
\ , 
\end{equation}
where $N^\perp = f = \left( -M + \frac{r^2}{\ell^2} + \frac{J^2}{4r^2}
\right)^{1/2}$, $N^\phi = - \frac{J}{2r^2}$, $\ell$ is a positive
parameter that sets the $\text{AdS}_3$ curvature scale, $\phi$ has
period~$2\pi$, and a non-extremal black hole is obtained when the mass
parameter $M$ and the angular momentum parameter $J$ satisfy $|J| <
M\ell$. The spacetime has many similarities with the Kerr black hole,
but its null infinities are asymptotically $\text{AdS}$, as opposed to
asymptotically flat. The conformal diagram of the $J=0$ case is shown
in Figure~\ref{fig:BTZstatic}. The importance of this asymptotic
structure for us is that the spacetime is not globally hyperbolic, and
to build a sensible quantum field theory one must impose boundary conditions at the
infinity. We shall see that the detector response turns out to be highly
sensitive to these boundary conditions.

We consider a massless, conformally coupled field.  We first introduce
on the covering space $\text{AdS}_3$ the three $\text{AdS}$-invariant
states whose Wightman functions are given by \cite{Carlip:1995}
\begin{equation}
\label{eq:AdSWightman}
G^{(\zeta)}_{A}(\x,\x')=\frac{1}{4\pi} \! 
\left(\frac{1}{\sqrt{{\Delta\X}^2(\x,\x')}}
-\frac{\zeta}{\sqrt{{\Delta\X}^2(\x,\x')+4\ell^2}}\right),
\end{equation}
where $\Delta\X^2(\x,\x')$ is the squared geodesic distance between
$\x$ and $\x'$ in the flat $\mathbb{R}^{2,2}$ in which $\text{AdS}_3$
can be embedded as a submanifold, the parameter $\zeta\in\{0,1,-1\}$
specifies whether the boundary condition at infinity is respectively
transparent, Dirichlet or Neumann, and we have suppressed the
$i\epsilon$ that controls the short distance
form~\eqref{eq:3dHadamard}. The Wightman function in the induced state
on the BTZ spacetime is then given by the image sum~\cite{Carlip:1995}
\begin{equation}
\label{eq:BTZWightmanSum}
{\textstyle G_{\text{BTZ}}(\x,\x')=\sum_n\,G_A(\x,\Lambda^n \x')} 
\ , 
\end{equation}
where $\Lambda \x'$ denotes the action on $\x'$ of the isometry
$(t,r,\phi)\mapsto (t,r,\phi+2\pi)$, and the notation suppresses the
distinction between points on $\text{AdS}_3$ and points on the
quotient spacetime.  The scalar field is assumed untwisted so that no
additional phase factors appear in~\eqref{eq:BTZWightmanSum}.

The detector's transition rate is obtained by substituting
\eqref{eq:BTZWightmanSum} into~\eqref{eq:resp3d:sharp:rate}.  
In sections \ref{sec:corot} and \ref{sec:inertial}
we discuss the transition rate for selected 
detector trajectories.

\section{Co-rotating detector in BTZ}
\label{sec:corot}

As the first example we consider 
a detector that is in the exterior region of the BTZ black
hole, at constant value of $r$ and co-rotating with the horizon
angular velocity~$\Omega_H$. In the special case $J=0$, we have
$\Omega_H=0$ and the detector is static. Unlike in Kerr, these
detector trajectories exist at arbitrarily large values of~$r$: 
this is a consequence of the AdS asymptotics. 

As the detector is stationary, we take the switch-on to be in the
asymptotic past. The Wightman function turns out to contain
singularities between timelike-separated points on the detector's
trajectory, but the consequent singularities in the transition rate
formula \eqref{eq:resp3d:sharp:rate} are integrable and the transition
rate remains well defined. Further, contour manipulations allow the
transition rate to be cast in a manifestly nonsingular form that is
amenable to analytic techniques, including asymptotic analyses in a
number of asymptotic regimes, as well as to numerical evaluation. We
can in particular verify analytically that the transition rate
satisfies
\begin{equation}
\dot{\mathcal{F}}(E) = \mathrm{e}^{-E/T_{\text{loc}}}\dot{\mathcal{F}}(-E) ,
\label{eq:KMS-Tloc}
\end{equation}
where $T_{\text{loc}}$ is the co-rotating Hawking temperature at the
detector's location~\cite{Carlip:1995}. 
(As the transition rate is stationary, 
we have dropped the subscript~$\tau$.) 
The transition rate is hence thermal in the
local Hawking temperature in the sense of the Kubo-Martin-Schwinger (KMS) 
property~\cite{Kubo:1957mj,Martin:1959jp}, 
as expected from the general properties of the 
Hartle-Hawking vacuum~\cite{Hartle:1976tp,Israel:1976ur}. 

\begin{figure}[t!]  
\centering
\subfloat[$\zeta=0$]{\includegraphics[width=0.3\textwidth]{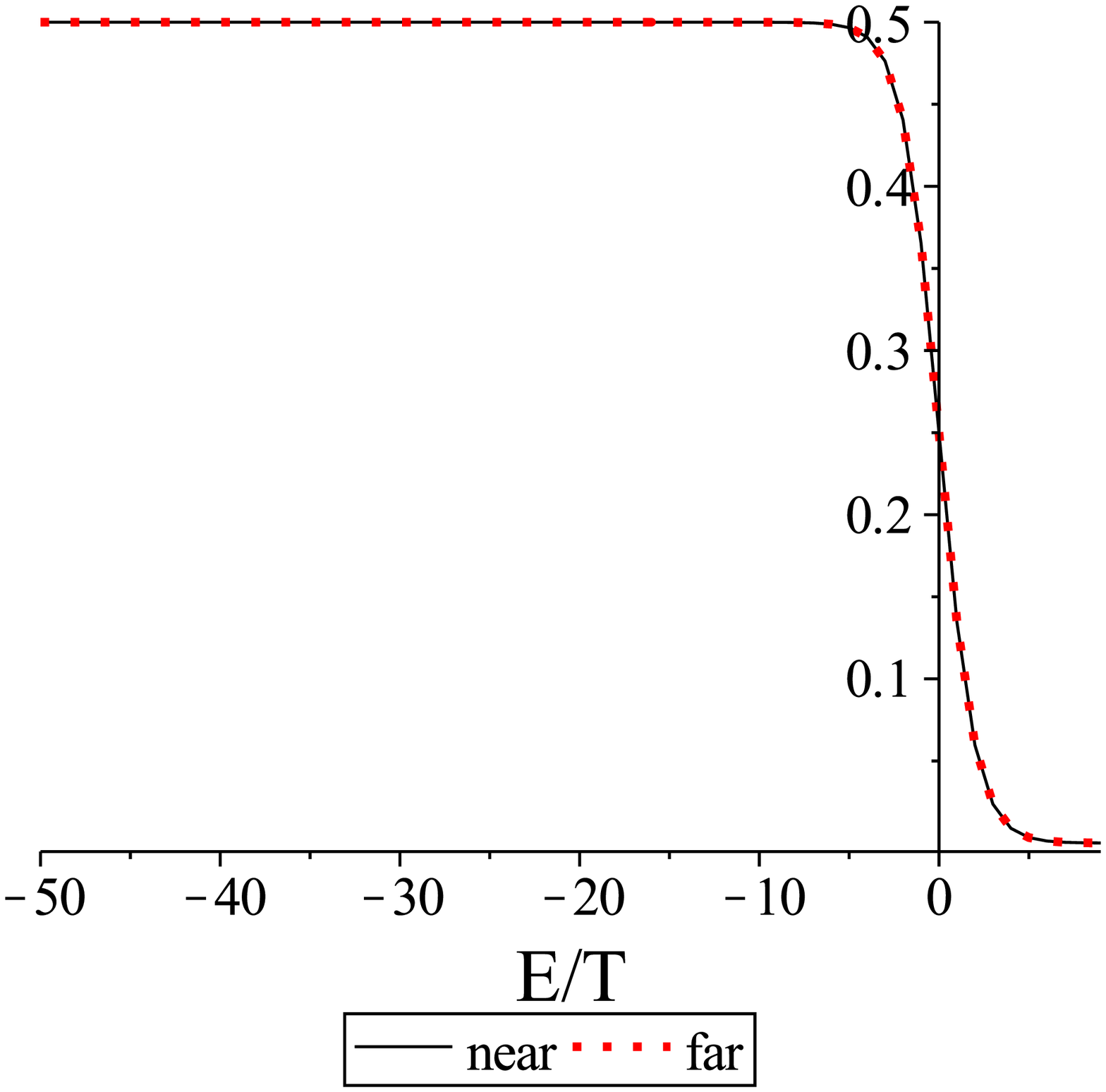}}            
\subfloat[$\zeta=1$]{\includegraphics[width=0.3\textwidth]{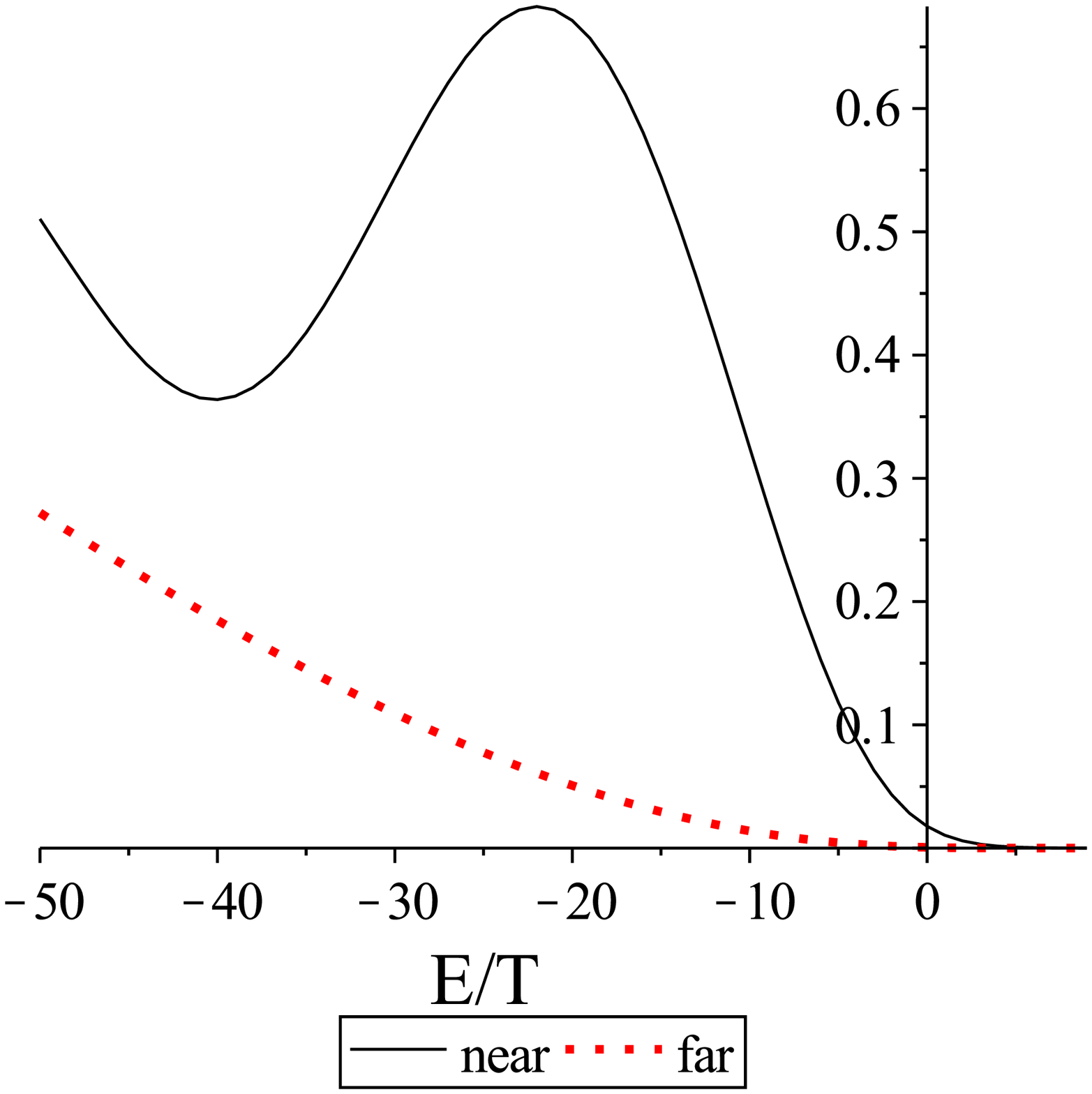}}  
\subfloat[$\zeta=-1$]{\includegraphics[width=0.3\textwidth]{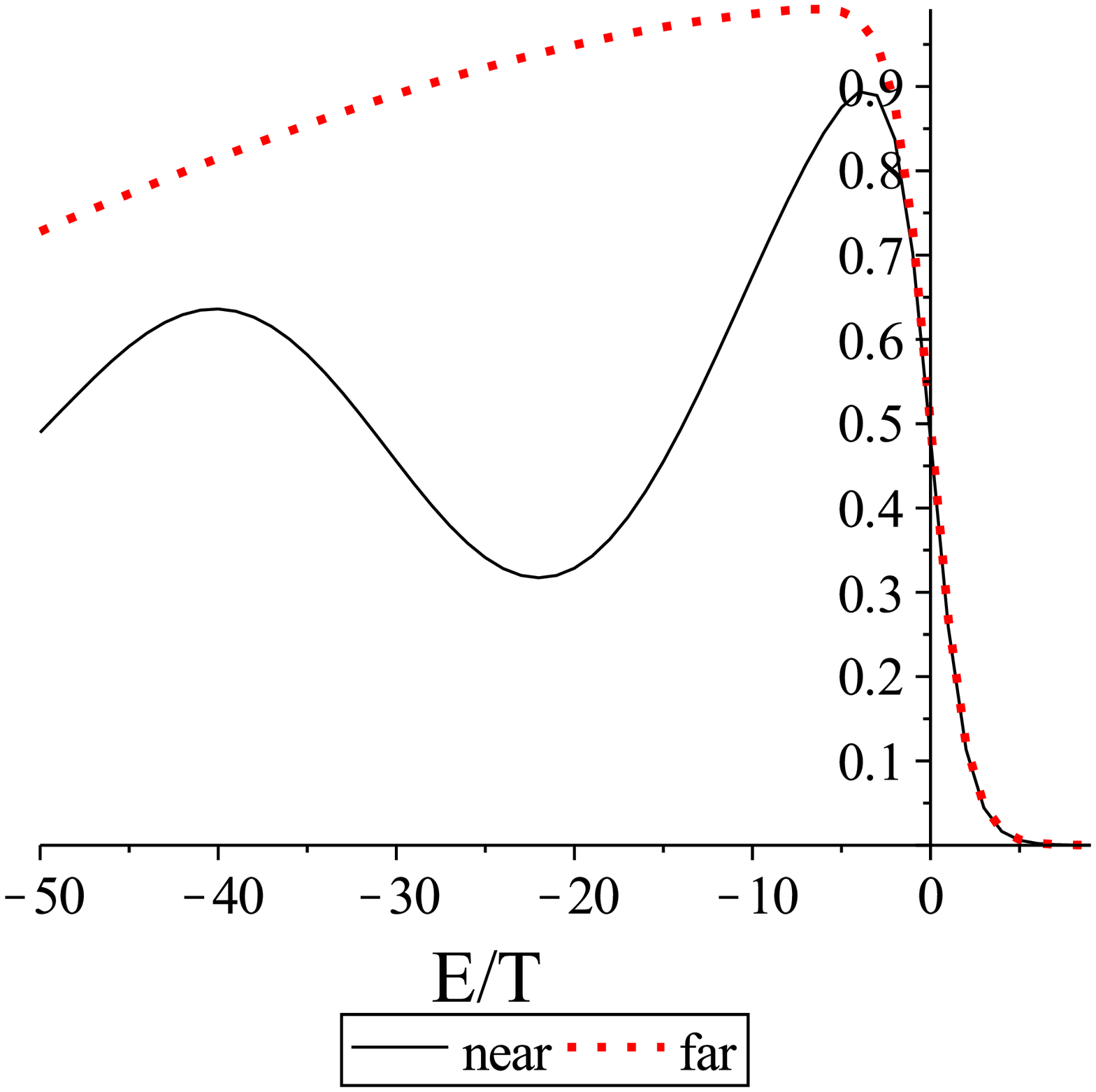}}  
\caption{$\dot{\mathcal{F}}$ for a co-rotating detector, 
as a function of the detector's energy
gap $E$ divided by the local Hawking temperature~$T$, for 
a large non-spinning hole, with the detector near the hole (solid) and far
from hole (dotted). Note the significant differences between the three
boundary conditions. 
\label{fig:transrate_rp10}} 
\end{figure}

The boundary condition
at the infinity is found to have a significant effect on the 
quantitative properties of the 
transition rate. The special case of a spinless black hole, with a
detector at large and small distances from the hole, is illustrated in
Figure~\ref{fig:transrate_rp10}.

\section{Inertial detector in BTZ}
\label{sec:inertial}

As the second example we consider a detector on a geodesic that falls radially into 
the spinless black hole. This trajectory is not stationary and the transition 
rate depends on both the switch-on moment and the switch-off moment. Furthermore the switch-on moment 
cannot be pushed to the infinite past because the trajectory starts at
the white hole singularity at a finite proper time. 

We have found no parameter ranges where the transition rate would be thermal 
in the sense of the KMS property~\eqref{eq:KMS-Tloc}. One situation where approximate thermality 
might have been expected is near the turning point of a trajectory far from the horizon. 
However, in this case the transition rate just reduces to 
that of a geodesic detector in $\text{AdS}_3$, which can be verified not to satisfy the KMS property. 
These observations are compatible with embedding space arguments which suggest that a
detector in $\text{AdS}_3$ should respond thermally only when its proper acceleration
exceeds $1/\ell$ \cite{Deser:1997ri,Deser:1998bb,Deser:1998xb,Russo:2008gb}. 

\begin{figure}[h!]
\centering
\includegraphics[scale=0.9]{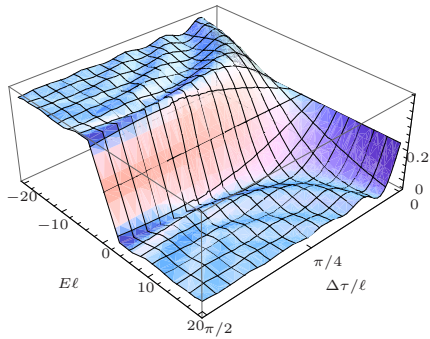}                
\caption{The transition rate of a detector on a radial geodesic in the spinless BTZ spacetime, assuming that the mass is large and that the switch-on and switch-off moments are not close to the white hole and black hole singularities, 
with the transparent boundary condition at the infinity. The horizontal axes are the detector's energy gap 
$E$ and the total detection time 
$\Delta\tau:=\tau-\tau_0$, normalised by the AdS scale~$\ell$. 
Note the dominance of the de-excitation rate ($E<0$) over the excitation 
rate ($E>0$) after the transient switch-on effects have died out.} 
\label{fig:zeroth3DBC0}
\end{figure}

We were however able to analyse the transition rate 
by a combination of asymptotic methods and numerical methods. Figure \ref{fig:zeroth3DBC0} shows a plot of 
the transition rate when the black hole is large and the switch-on and switch-off moments are not close to the 
white hole and black hole singularities, 
with the transparent boundary condition at the infinity. 

\section{Concluding remarks}

That the response of a co-rotating detector in the BTZ spacetime is thermal in the co-rotating Hawking temperature was to be expected from the general properties of the Hartle-Hawking vacuum~\cite{Hartle:1976tp,Israel:1976ur}. 
Our formalism allowed us analyse this thermal response quantitatively, by a combination of analytic and numerical techniques. 
We found in particular that the response depends strongly on the choice of the boundary condition at the infinity. 
We also showed perturbatively that the response loses its thermal character when the detector's 
angular velocity differs from that of the black hole. 

For a detector falling radially into a spinless BTZ hole, we found no parameter space regions where the 
transition rate would exhibit thermality. The transition rate is again affected by the choice of the boundary condition at the infinity, but this effect appears to be subdominant to those caused by the switching and the motion. 

It would be interesting to compare our results for the transition rate in the BTZ spacetime to that in 
Schwarzschild spacetime. For example, one may expect an inertial detector in Schwarzschild to respond to the 
Hartle-Hawking vacuum approximately thermally in the asymptotically flat region,
owing to the asymptotic flatness of Schwarzschild. We leave these questions subject to future work.

\section*{Acknowledgments}

J.~L. thanks the organisers of the ``Bits, Branes, Black Holes'' 
programme for hospitality at the 
Kavli Institute for Theoretical Physics,
University of California at Santa Barbara. 
This research was supported in part by the National 
Science Foundation under Grant No.\ NSF PHY11-25915 and by 
the Science and Technology Facilities Council\null.

\section*{References}

\bibliography{ae100prghod}

\begin{thebibliography}{10}

\bibitem{HBTZ}
Ba\~nados, M., Henneaux, M., Teitelboim, C.  and Zanelli, J., ``{Geometry of
  the (2+1) black hole}'', {\em Phys. Rev}, {\bf D48}, 1506--1525, (1993).
  {\small[\href{http://arxiv.org/abs/gr-qc/9302012}{{arXiv:gr-qc/9302012
  {\small[gr-qc]}}}]}.

\bibitem{BTZ}
Ba\~nados, M., Teitelboim, C.  and Zanelli, J., ``{The black hole in
  three-dimensional space-time}'', {\em Phys. Rev. Lett.}, {\bf 69},
  1849--1851, (1992).
  {\small[\href{http://arxiv.org/abs/hep-th/9204099}{{arXiv:hep-th/9204099
  {\small[hep-th]}}}]}.

\bibitem{Carlip:1995}
Carlip, S., ``{The (2+1)-dimensional black hole}'', {\em Class. Quant. Grav.},
  {\bf 12}, 2853--2880, (1995).
  {\small[\href{http://arxiv.org/abs/gr-qc/9506079}{{arXiv:gr-qc/9506079
  {\small[gr-qc]}}}]}.

\bibitem{Decanini:2005gt}
Decanini, Y.  and Folacci, A., ``{Off-diagonal coefficients of the
  Dewitt-Schwinger and Hadamard representations of the Feynman propagator}'',
  {\em Phys. Rev.}, {\bf D73}, 044027, (2006).
  {\small[\href{http://arxiv.org/abs/gr-qc/0511115}{{arXiv:gr-qc/0511115
  {\small[gr-qc]}}}]}.

\bibitem{Deser:1997ri}
Deser, S.  and Levin, O., ``{Accelerated detectors and temperature in
  (anti)-de~Sitter spaces}'', {\em Class. Quant. Grav.}, {\bf 14}, L163--L168,
  (1997).
  {\small[\href{http://arxiv.org/abs/gr-qc/9706018}{{arXiv:gr-qc/9706018
  {\small[gr-qc]}}}]}.

\bibitem{Deser:1998bb}
Deser, S.  and Levin, O., ``{Equivalence of Hawking and Unruh temperatures
  through flat space embeddings}'', {\em Class. Quant. Grav.}, {\bf 15},
  L85--L87, (1998).
  {\small[\href{http://arxiv.org/abs/hep-th/9806223}{{arXiv:hep-th/9806223
  {\small[hep-th]}}}]}.

\bibitem{Deser:1998xb}
Deser, S.  and Levin, O., ``{Mapping Hawking into Unruh thermal properties}'',
  {\em Phys.Rev.}, {\bf D59}, 064004, (1999).
  {\small[\href{http://arxiv.org/abs/hep-th/9809159}{{arXiv:hep-th/9809159
  {\small[hep-th]}}}]}.

\bibitem{deWitt}
DeWitt, B.~S., ``Quantum gravity: the new synthesis'', in Hawking, S.~W.  and
  Israel, W., eds., {\em General Relativity; an Einstein centenary survey}, p.
  680, (Cambridge University Press, Cambridge; New York, 1979).

\bibitem{Fewster:1999gj}
Fewster, C.~J., ``{A general worldline quantum inequality}'', {\em Class.
  Quant. Grav.}, {\bf 17}, 1897--1911, (2000).
  {\small[\href{http://arxiv.org/abs/gr-qc/9910060}{{arXiv:gr-qc/9910060
  {\small[gr-qc]}}}]}.

\bibitem{Hartle:1976tp}
Hartle, J.~B.  and Hawking, S.~W., ``{Path integral derivation of black hole
  radiance}'', {\em Phys. Rev.}, {\bf D13}, 2188--2203, (1976).

\bibitem{hodgkinson-louko:beyond4d}
Hodgkinson, L.  and Louko, J., ``{How often does the Unruh-DeWitt detector
  click beyond four dimensions?}'', {\em J. Math. Phys.}, {\bf 53}, 082301,
  (2012). {\small[\href{http://arxiv.org/abs/1109.4377}{{arXiv:1109.4377
  {\small[gr-qc]}}}]}.

\bibitem{Hodgkinson:2012mr}
Hodgkinson, L.  and Louko, J., ``{Static, stationary and inertial Unruh-DeWitt
  detectors on the BTZ black hole}'', {\em Phys. Rev.}, {\bf D}, to appear,
  (2012). {\small[\href{http://arxiv.org/abs/1206.2055}{{arXiv:1206.2055
  {\small[gr-qc]}}}]}.

\bibitem{Israel:1976ur}
Israel, W., ``{Thermo-field dynamics of black holes}'', {\em Phys. Lett.}, {\bf
  A57}, 107--110, (1976).

\bibitem{junker}
Junker, W.  and Schrohe, E., ``{Adiabatic vacuum states on general space-time
  manifolds: Definition, construction, and physical properties}'', {\em Annales
  Henri Poincar\'e Phys. Theor.}, {\bf 3}, 1113--1182, (2002).
  {\small[\href{http://arxiv.org/abs/math-ph/0109010}{{arXiv:math-ph/0109010
  {\small[math-ph]}}}]}.

\bibitem{kay-wald}
Kay, B.~S.  and Wald, R.~M., ``{Theorems on the uniqueness and thermal
  properties of stationary, nonsingular, quasifree states on space-times with a
  bifurcate Killing horizon}'', {\em Phys. Rept.}, {\bf 207}, 49--136, (1991).

\bibitem{Kubo:1957mj}
Kubo, R., ``{Statistical mechanical theory of irreversible processes. 1.
  General theory and simple applications in magnetic and conduction
  problems}'', {\em J. Phys. Soc. Jap.}, {\bf 12}, 570--586, (1957).

\bibitem{Langlois}
Langlois, P., ``{Causal particle detectors and topology}'', {\em Annals Phys.},
  {\bf 321}, 2027--2070, (2006).
  {\small[\href{http://arxiv.org/abs/gr-qc/0510049}{{arXiv:gr-qc/0510049
  {\small[gr-qc]}}}]}.

\bibitem{louko-satz:profile}
Louko, J.  and Satz, A., ``{How often does the Unruh-DeWitt detector click?
  Regularisation by a spatial profile}'', {\em Class. Quant. Grav.}, {\bf 23},
  6321--6344, (2006).
  {\small[\href{http://arxiv.org/abs/gr-qc/0606067}{{arXiv:gr-qc/0606067
  {\small[gr-qc]}}}]}.

\bibitem{satz-louko:curved}
Louko, J.  and Satz, A., ``{Transition rate of the Unruh-DeWitt detector in
  curved spacetime}'', {\em Class. Quant. Grav.}, {\bf 25}, 055012, (2008).
  {\small[\href{http://arxiv.org/abs/0710.5671}{{arXiv:0710.5671
  {\small[gr-qc]}}}]}.

\bibitem{Martin:1959jp}
Martin, P.~C.  and Schwinger, J.~S., ``{Theory of many-particle systems. 1.}'',
  {\em Phys. Rev.}, {\bf 115}, 1342--1373, (1959).

\bibitem{Russo:2008gb}
Russo, J.~G.  and Townsend, P.~K., ``{Accelerating branes and brane
  temperature}'', {\em Class. Quant. Grav.}, {\bf 25}, 175017, (2008).
  {\small[\href{http://arxiv.org/abs/0805.3488}{{arXiv:0805.3488
  {\small[hep-th]}}}]}.

\bibitem{satz:smooth}
Satz, A., ``{Then again, how often does the Unruh-DeWitt detector click if we
  switch it carefully?}'', {\em Class. Quant. Grav.}, {\bf 24}, 1719--1732,
  (2007).
  {\small[\href{http://arxiv.org/abs/gr-qc/0611067}{{arXiv:gr-qc/0611067
  {\small[gr-qc]}}}]}.

\bibitem{schlicht}
Schlicht, S., ``{Considerations on the Unruh effect: Causality and
  regularization}'', {\em Class. Quant. Grav.}, {\bf 21}, 4647--4660, (2004).
  {\small[\href{http://arxiv.org/abs/gr-qc/0306022}{{arXiv:gr-qc/0306022
  {\small[gr-qc]}}}]}.

\bibitem{Sriramkumar:1994pb}
Sriramkumar, L.  and Padmanabhan, T., ``{Response of finite time particle
  detectors in noninertial frames and curved space-time}'', {\em Class. Quant.
  Grav.}, {\bf 13}, 2061--2079, (1996).
  {\small[\href{http://arxiv.org/abs/gr-qc/9408037}{{arXiv:gr-qc/9408037
  {\small[gr-qc]}}}]}.

\bibitem{unruh}
Unruh, W.~G., ``{Notes on black hole evaporation}'', {\em Phys. Rev.}, {\bf
  D14}, 870--892, (1976).

\end{thebibliography}

\end{document}